# Nucleosome positioning: resources and tools online

Vladimir B. Teif

School of Biological Sciences, University of Essex, Wivenhoe Park, Colchester CO4 3SQ, UK
E-mail: vteif@essex.ac.uk or v.teif@dkfz-heidelberg.de

**Abstract**

Nucleosome positioning is an important process required for proper genome packing and its accessibility to execute the genetic program in a cell-specific, timely manner. In the recent years hundreds of papers have been devoted to the bioinformatics, physics and biology of nucleosome positioning. The purpose of this review is to cover a practical aspect of this field, namely to provide a guide to the multitude of nucleosome positioning resources available online. These include almost 300 experimental datasets of genome-wide nucleosome occupancy profiles determined in different cell types and more than 40 computational tools for the analysis of experimental nucleosome positioning data and prediction of intrinsic nucleosome formation probabilities from the DNA sequence. A manually curated, up to date list of these resources will be maintained at http://generegulation.info.

**Introduction**

The nucleosome is the basic unit of chromatin compaction, composed of the histone octamer and 146-147 base pairs (bp) of DNA wrapped around it. Nucleosomes can form at any genomic locations, but some DNA sequences have higher affinity to the histone octamer, mostly due to the differential bending properties of the DNA double helix. In addition, nucleosome positioning is cell type-specific, in a sense that the cells of the same organism sharing the same genome can have different nucleosome locations depending on the cell type and state. Interested readers are directed to a number of recent publications reviewing the biological, physical and bioinformatics aspects of these phenomena, which will be outside of the scope of the current work [1-32]. Here we will omit fundamental scientific questions, and will focus on a very practical aspect of the field: which experimental nucleosome positioning datasets already exist, how to generate your own data, and how to compare these with other experimental datasets and bioinformatically predicted nucleosome positions in a given genome?

**1. Available experimental datasets**

Recent high-throughput genome-wide data with respect to nucleosome positioning come from a number of related techniques, which have in common an idea to cut DNA between nucleosomes and map protected DNA regions. The most frequently used method is MNase-seq (chromatin digestion by micrococcal nuclease followed by deep sequencing) [11, 33-35]. A number of complementary methods have been proposed using MNase alone or in combination with sonication [36] and/or histone H3 immunoprecipitation (ChIP-seq) or other enzymes such as DNase (DNase-seq) [37, 38], transposase (ATAC-seq) [39, 40] and CpG methyltransferase (NOME-seq) [41]. Another possibility is to use directed chemical cleavage, either by hydroxyl radicals targeted by artificially introduced histone modifications [42-44], or by small aromatic molecules such as methidiumpropyl-EDTA which preferentially intercalate in the DNA double helix between nucleosomes and cleave it in the presence of Fe(II) ions (MRE-seq) [45]. The method of tiling microarrays (ChIP-chip), which several years ago was the method of choice [46-49], is currently mostly overridden by high throughput sequencing based methods. New methods continue to appear to address three major problems: (1) DNA sequence biases of the experimental setup; (2) targeted enrichment to achieve ultra-high sequencing coverage for a subset of genomic regions of interest; (3) single-cell sequencing to overcome the problem of heterogeneity, which is present in any cell population.

Unlike typical transcription factor (TF) or histone-modification ChIP-seq datasets, nucleosome positioning aims to be determined with a single-base pair resolution, which requires higher sequencing coverage. This distinction becomes particularly important for large genomes. Therefore, while Yeast nucleosome positioning datasets could be still stored even at individual web sites [10, 50-54], for higher eukaryotes the question of the storage and location of such datasets becomes outstanding. For example, the first investigation of genome-wide nucleosome redistribution in human cells was performed in 2008 [55]. The latter study reported nucleosome maps in activated versus resting $CD^{4+}$ T cells, determined with 10 base pair resolution using MNase-seq. Several studies later used this method for human cell lines [41, 56-62] and primary human cells [63-68], each generating about 200-400 million DNA reads for a single condition. However, a direct comparison of nucleosome positions in primary cells from patients vs. healthy individuals still remains a challenge because a definitive resolution of nucleosome positions for the human genome requires up to the order of 1-4 billion reads [69]. (For comparison, a typical ChIP-seq dataset addressing TF binding or histone modifications contains just several tens millions reads). This increases the required storage space to the order of terabytes instead of gigabytes per dataset as was common for the "first wave" of Next Generation Sequencing (NGS) experiments. Typically, the Short Read Archive (SRA) is used for such data storage nowadays [70], while a detailed documentation and processed files are stored in the Gene Expression Omnibus (GEO) database [71]. The problem is that usually it is difficult to know which datasets already exist for nucleosome positioning in a given cell type, because "nucleosome positioning" is not a specific type of experiment, and searching e.g. for the word "MNase-seq" returns only part of MNase-seq-type datasets. For example, a surprisingly large number of 14 datasets from different laboratories already exist for nucleosome positioning in a single cell type, mouse embryonic stem cells (ESCs) [45, 67, 72-79]. One would not be aware of all of them without tracking the corresponding publications. This was one of the motivations to create a complete list of such datasets.

Table 1 provides a summary of nucleosome positioning datasets which are currently available. The short one-line descriptions introduced here for each dataset do not substitute for more detailed GEO entries and have been used in this format specifically to help readers quickly navigate in the sea of nucleosomes. Historically, studies in *Yeast* have founded this type of research, and these, together with other popular model organisms such as *Arabidopsis* and *Drosophila* still provide valuable systems to understand fundamental biological mechanisms, especially when genetic modifications need to be introduced quickly. However, current literature is becoming highly populated by studies in mouse and human, since the latter have direct links to cancer and other diseases. In practical terms, knowing which datasets already exist can be helpful either to integrate their analysis with other NGS datasets available for the same cell type/state, or at least to get some hints about new experiment design. Visitors of the web site generegulation.info, where this list was initiated in 2009, frequently asked the following question: Suppose, we have no experimentally measured nucleosome maps for the cell type of our interest. Can we take nucleosome positions from another cell type (of the same organism) and compare these with our RNA-seq, ChIP-seq, etc? In general, the answer is "no", since nucleosome maps are cell-type specific. One has to determine nucleosome maps in a given cell type, which leads us to the next practical questions.

**Table 1. Experimental nucleosome positioning datasets sorted by cell type, newest first**

| Description | Accession # |
| --- | --- |
| *Human – about 50 datasets with different cell types/conditions:* | |
| HuRef lymphoblastoid line, α-satellite arrays of centromeres [58]. ChIP-seq. | GSE60951 |
| H1-OGN embryonic stem cells, H1-OGN induced pluripotent stem cells, and fibroblasts differentiated from H1-OGN ESCs [67]. MNase-seq. | GSE59062 |
| HCT116 colon cancer cells and their genetic derivatives which lack DNA methyltransferases DNMT3B and DNMT1 activity [57]. NOME-seq. | GSE58638 |
| Primary human endothelial cells stimulated with tumour necrosis factor alpha (TNFalpha) [63]. MNase-seq. | GSE53343 |
| MCF-7 (breast cancer) with and without MBD3 knockdown [59]. MNase-seq. | GSE51097 |
| Human embryonic stem cells (H1 and H9 hESCs). MNase-seq | GSE49140 |
| Human sperm [66]; Limited regions retain nucleosomes in sperm. MNase-seq. | GSE47843 |
| Human colo829 cell line. MNase-seq | GSE47802 |
| Raji cells (lymphoblastoid-like) with and without *α*-amanitin [60]. MNase-seq. | GSE38563 |
| 7 lymphoblastoid cell lines from the HapMap project [69]. MNase-seq. | GSE36979 |
| Lymphoblastoid GM12878 and K562 cell lines [56]. MNase-seq. | GSE35586 |
| CD36+ cells with and without BRG1 knockdown [68]. MNase-seq, ChIP-seq. | GSE26501 |
| Human embryonic carcinoma (NCCIT) cell line [61]. MNase-seq, ChIP-seq. | GSE25882 |
| Primary $CD^{4+}$ T-cells, $CD^{8+}$ T-cells and granulocytes [65]. MNase-seq. | GSE25133 |
| MCF7EcoR cells where P53 was either activated or not [62]. MNase-seq. | GSE22783 |
| Nucleosome positioning and DNA methylation in IMR90 [41]. NOME-seq. | GSE21823 |
| Resting and activated $CD^{4+}$ T cells [55]. MNase-seq; H3, H2A.Z ChIP-seq. | SRA000234 |
| *Mouse – about 70 datasets with different cell types/conditions:* | |
| Mouse ESCs [45]. MNase-seq, MPE-seq, MPE-ChIP-seq | GSE69098 |
| Mouse ESCs, wild type (WT) and Dnmt1/3a/3b triple knockout [78]. MNase-seq | GSE64910 |
| Mouse EScs, WT and remodeler BAF250a knockout [80]. MNase-seq. | GSE59082 |

| Mouse ESCs, induced pluripotent stem cells (iPCs), somatic tail-tip fibroblasts (TTF) and liver [67]. MNase-seq. | GSE59062 |
|---|---|
| Mouse ESCs and sperm [77]. Different size-selection of MNase-seq fragments. | GSE58101 |
| Mouse ESCs, siRNA knockdown of EGFP, Smarca4 or MBD3 [79]. MNase-seq | GSE57170 |
| Mouse ESCs, low MNase digestion; dinucleosome fraction [76]. MNase-seq. | GSE56938 |
| Mouse ESCs and differentiated iMEFs. RED-seq; paper not published yet. | GSE51821 |
| Mouse ESCs (J1) [72]. MNase-seq, ChIP-seq | GSE51766 |
| Mouse ESCs, low MNase digestion [81]. MNase-seq. | GSE50706 |
| Mouse ESCs (E14) and SMARCAD1-knock down cells. MNase-seq. | GSE47802 |
| Mouse ESCs and induced pluripotent cells (iPSC) from different layers [82] | GSE46716 |
| Mouse ESCs, neural progenitor cells (NPCs) and neurons with and without HMGN1 knockout. MNase-seq using high and low MNase digestion levels [73]. | GSE44175 |
| Mouse ESCs, NPCs and embryonic fibroblasts (MEFs) [75]. MNase-seq. | GSE40951 |
| Mouse thymocytes, MNase-seq. | GSE69474 |
| Mouse B-cell to macrophage lineage switching, several time points. MNase-seq. | GSE53460 |
| Mouse liver, 3-mohth and 21-month old mice [83]. MNase-seq. | GSE58005 |
| Mouse liver, 6 time points of the 24h light:dark cycle; WT and Bmal1-/- [84]. | GSE47142 |
| Mouse liver [74]. MNase-seq and ChIP-seq. | GSE26729 |
| Mouse bone marrow-derived macrophages (BMDMs) [85]. MNase-seq. | GSE62151 |
| Hypothalamus from MeCP2 knockout mice and control mice [86]. MNase-seq. | GSE66869 |
| Cultured germline stem cells with and without Scml2 knockout [87]. MNase-seq. | GSE55060 |
| Primary CD4+ CD8+ DP thymocytes and Rag2 -/- thymocytes [88]. MNase-seq. | GSE56395 |
| Fibroblasts from E13.5 embryos. WT, Snf5-/- and Brg1-/- [89]. MNase-seq. | GSE38670 |
| *Drosophila melanogaster, MNase-seq:* | |
| S2 cell line. WT and stimulated by heat killed Salmonella typhimurium. | GSE64507 |
| S2 cell line. WT; treated with RNAi against Beta-galactosidase or GAGA [90]. | GSE58957 |
| S2 cell line. WT and Beaf32-depleted [91]. | GSE57166 |
| S2 cell line. WT and depletion of CTCF/P190 and ISWI [92]. | GSE51599 |
| S2 cell line, WT [93]. | GSE49526 |
| Staged Drosophila embryos [94]. | GSE41686 |
| S2 cell line. WT, mock-treated, and NELF-depleted [95]. | GSE22119 |
| *Arabidopsis thaliana:* | |
| Col-0 seeds; chr11-1 chr17-1, MNase-seq [96] | GSE50242 |
| Col-0 seeds; WT and inhibition of Pol V-produced lncRNAs. MNase-seq [97] | GSE38401 |
| Col-0 seeds, shoots; MNase-seq, ChIP-seq, Bisulfite sequencing [98] | GSE21673 |
| *Caenorhabditis elegans, MNase-seq:* | |
| Mixed stage, wild-type (N2) C. elegans. SOLiD paired-end sequencing [99] | SRX000426 |
| *Chlamydomonas reinhardtii:* | |
| Chlamydomonas strain CC 1609. MNase-seq [100] | GSE62690 |
| *Saccharomyces cerevisiae and related species, MNase-seq:* | |
| *S. cerevisiae* hho1, ioc3isw1, and chd1 deletion mutants complemented with the corresponding copies from *K. lactis* [101]. | GSE66979 |
| *S. cerevisiae*. Strain W303, stationary growth phase. Wild type (WT) and with introduced DNMT3b [102]. | GSE66907 |
| *S. cerevisiae*. Strains carrying the Sth1 degron allele and either pGal-UBR1 (YBC3386) or ubr1 null (YBC3387) represent RSC null and RSC wild type correspondingly [103]. | GSE65593 |
| *S. cerevisiae*. WT and Snf2 K1493R, K1497R strains; unstressed/stressed [104] | GSE61210 |
| *S. cerevisiae*. Strain W303. WT and modification affecting one of the following chromatin remodelers: ISW1, CHD1, FUN30, IOC3 [43]. | GSE59523 |
| *S. cerevisiae*. Strain W303. Affected histone deacetylases Sir2 and Rpd3 [105]. | GSE57618 |
| *S. cerevisiae*. Strain YK699, WT and changes addressing the following: Scc2-4; Sth1-3; a2/MCM1; TATAC; TATAΔ. Replicates at 25°C and 37°C [106]. | GSE56994 |
| *S. cerevisiae*. Calorie restricted and non-restricted WT, ISW2DEL and ISW2K215R strains [107] | GSE53718 |
| *S. cerevisiae*. Strain W303 (yFR212) [108]. MNase-seq and H2A.Z ChIP-seq | GSE47073 |
| *S. cerevisiae*. Strain S288c (BY4741). "Young yeast", "old yeast", and "old yeast with histone over expression" [109]. | GSE47023 |
| *S. cerevisiae*. Strain BY4741, WT and Hog1 mutant. Exposed/not exposed to osmostress [110]. | GSE41494 |
| *S. cerevisiae*. Strain BY4742, WT, Ssn6 KO and Tup1 KO [111]. | GSE37465 |
| *S. cerevisiae*. Strain S288C. WT, Nup170Δ and Sth1p depletion [112]. | GSE36792 |
| *S. cerevisiae*. Strain BY4741. Study of response to $H_2O_2$ over time in the S288c derivative [113]. | GSE30900 |
| *S. cerevisiae*. Strain YEF473A. WT and mutant with H3 shutoff to study histone H3 depletion [114]. | GSE29292 |
| WT and mutant strains in *S. cerevisiae*, *C. albicans*, and *S. pombe* [115]. | GSE28839 |
| *S. cerevisiae* at varying phosphate concentrations | GSE26392 |
| *S. cerevisiae*. Strain XF218. H3 Chip-seq [116] | GSE23778 |
| 12 Ascomycete species: *Saccharomyces mikatae, Saccharomyces bayanus, Saccharomyces castellii, Saccharomyces cerevisiae, Kluyveromyces waltii, Saccharomyces paradoxus, Candida glabrata, Candida albicans, Debaryomyces hansenii, Kluyveromyces lactis, Saccharomyces kluyveryii, Yarrowia lipolytica* [117]. | GSE22211 |
| Comparison of nucleosome positioning in *S. cerevisiae, S. paradoxus* and their hybrid for wild-type and deletion mutant strains [118]. | GSE18939 |
| *S. cerevisiae*. Strains BY4741 and RPO21. MNase titration series from three different titration levels – underdigested, typical digestion, and overdigested BY4741 cells. Time dependence series: MNase-seq at 0, 20, and 120 minutes after shifting RPO21 cells from 25 C to 37 C [34]. | GSE18530 |
| *S. cerevisiae*. Chromatin remodelling by Isw2 [50]. Tiling microarrays. http://research.fhcrc.org/tsukiyama/en/genomics-data/global_nucleosomemapping.html | GSE8813, GSE8814, GSE8815 |
| The Penn State Genome Cartography Project. *S. cerevisiae* and *D. melanogaster* [10, 53, 54, 119]. Tiling microarrays. http://atlas.bx.psu.edu | |
| *Saccharomyces pombe, MNase-seq:* | |
| Strain Hu1867. WT and without Fun30 chromatin remodeler Fft3 [120]. | GSE58012 |
| Strain FWP172. WT and spt6-1 at two different MNase concentrations [121]. Spt6 is a histone chaperone. | GSE49572 |
| Wild type and without CHD remodeler Hrp3 [122]. | GSE40451 |
| Strain D18, log phase and stationary Phase [123]. http://www.acsu.buffalo.edu/~mjbuck/Fission_Yeast_chromatin.html | GSE28071 |

## 2. Computational tools to analyze nucleosome positioning data

Many laboratories nowadays consider nucleosome positioning as an important additional piece of information to supplement new stories about a specific TF, pathway, or biological process happening in chromatin. If a general workflow of NGS analysis is already established in the lab, it is tempting to consider nucleosome positioning as just another dataset. Indeed, the first steps of analysis of nucleosome positioning experiments (e.g., MNase-seq, histone H3 ChIP-seq, NOME-seq), require standard NGS software for read mapping and quality control. However, next steps, which use as input mapped BED/BAM/SAM files, diverge from the analysis of typical ChIP-seq experiments. The major difference is that in the nucleosome positioning field we are dealing with millions of small, fuzzy enrichment peaks corresponding to individual nucleosomes defined with precision of one to several bp, in contrast to TF ChIP-seq where one deals with better defined sharp peaks, or histone-modifications ChIP-seq where one deals with broad peaks determined at a precision of hundreds bp. Therefore, generic peak calling programs usually used for ChIP-seq, such as MACs [124] or HOMER [125], are not optimal for nucleosome position calling. Yet, the basic idea behind nucleosome position calls from the experimental data is the same: one has to detect enriched peaks of size around 147 bp (e.g. TemplateFilter [34], NPC [126], nucleR [127], NOrMAL [128], PING/PING2 [108, 129], MLM [46], NucDe [130], NucleoFinder [131], ChIPseqR [132], NSeq [133], NucHunter [134], iNPS [135] and PuFFIN [136]). Alternatively, one does not call nucleosome positions at all, and instead operates with the continuous nucleosome occupancy profile, defining regions of cell type/state specific differential occupancy (e.g. DANPOS/DANPOS2 [137], DiNuP [138], NUCwave [139]). We have been also applying the latter idea when analysing nucleosome positioning in mouse and human [12, 26, 76] using custom made scripts (Vainshtein and Teif, unpublished). As a rule of thumb, calling nucleosome peaks is reasonable when these are well defined. This is typically the case in yeast but not in higher eukaryotes (unless a specific subset of well-positioned nucleosomes is considered). On the other hand, nucleosome landscapes in mouse and human are usually easier to interpret in terms of differential occupancy changes. An additional complication is that since typical MNase-seq experiments are being performed using 10,000-1,000,000 cells, the averaged nucleosome profile characterising this cellular ensemble does not represent any particular individual cell. Therefore, another analysis approach is to reconstruct the most probable non-overlapping nucleosome positions in individual cells using Monte Carlo simulations (e.g. NucPosSimulator [140]). In addition, a number of questions related to nucleosome positioning, such as e.g. enhancer identification, have been addressed in more specific tools [141]. Finally, biologically oriented users without solid knowledge of programming can take advantage of a number of available programs which visually display nucleosome occupancy in their region of interest (e.g. Skyline Nucleosome browser [142] or earlier nucleosome repositories visualizing yeast nucleosome maps [10, 50-54]). In principle, it is also possible to use generic genomic visualization tools such the UCSC Genome Browser [143]. However, due to large sizes of typical nucleosome positioning datasets this is not optimal. Thus, at least 21 computational tools offering one of these five options of the primary analysis of nucleosome positioning data are currently available online, listed with the corresponding explanations in Table 2. Since the analysis of nucleosome positioning performed using one of these five workflows (or combinations of them) can lead to equally interesting biological results, it is beyond the aims of the current review to make recommendations about the choice of these programs. Concerning the subclass of nucleosome peak calling software, interested readers are referred to recent reviews where some of the performance characteristics of nucleosome peak calling algorithms have been compared [139]. An additional parameter to be considered is the popularity of the software. Here, popularity is defined as the number of literature references in Google Scholar to the original paper. Of course, popularity does not necessarily reflect scientific superiority or the ease of use. Furthermore, since input/output data and the purpose of the software are very different, the superiority cannot be defined. The popularity is also not a very robust indicator because some newer tools might have fewer citations due to publication time lapse, while older items might have more citations because the original paper also introduced new experimental data / analysis. In addition, one has to take into account that highly cited software for the analysis of nucleosome positioning from tiling microarrays [46, 47] is not applicable to the nowadays experiments using NGS sequencing. With this disclaimer in mind, currently most popular nucleosome peak callers for single- or paired-end MNase-type nucleosome positioning experiments are TemplateFilter [34], NPC [126], DANPOS/DANPOS2 [137], nucleR [127], NOrMAL [128] and PING/PING2 [108, 129]. Paired-end sequencing is a more recent version of the experimental setup, and not all programs support it. Last but not the least feature is whether the program is a command-line tool, whether it is compatible with R or MATLAB, and whether it has a graphical user interface (GUI). These features are specified in Table 2.

**Table 2. Software to process NGS nucleosome experiments (sorted by popularity).**

| Description | Command line / R / MATLAB / JAVA GUI | Tiling arrays | Peak calling | Paired-end | # citations |
|---|---|---|---|---|---|
| **Tiling array analysis** [47]. A Matlab code, which is complemented by **MLM** and **NucleR** packages. http://bcb.dfci.harvard.edu/~gcyuan/software.html | -/-/+/- | + | + | - | 942 |
| **TemplateFilter**: Perl source code and executable files for nucleosome positioning data processing [34]. Applicable to Solexa-type high-throughput sequencing data. http://compbio.cs.huji.ac.il/NucPosition/TemplateFiltering/Home.html | +/-/-/- | - | + | - | 181 |
| **NPS**: Nucleosome Positioning from Sequencing [126]. This is Python based nucleosome peak caller, which is recommended for the use together with software **BINOCh** from the same group. http://liulab.dfci.harvard.edu/NPS/ | +/-/-/- | - | + | - | 95 |
| **DANPOS** and **DANPOS2**: Dynamic Analysis of Nucleosome Positioning and Occupancy by Sequencing [137]. This is a Python package, which reports changes in location, fuzziness, or occupancy for a given nucleosome or any genomic region. It allows generating aggregate profile plots and heatmaps for subsets of genomic regions. https://sites.google.com/site/danposdoc/ | +/-/-/- | - | + | + | 32 |
| **nucleR**: Non-parametric nucleosome positioning. This is an R package included in the Bioconductor [127]. It allows treating both NGS and Tiling Arrays experiments. The software is integrated with standard genomics R packages and allows for *in situ* visualization as well as to export results to common genome browser formats. http://mmb.pcb.ub.es/nucleR/ | -/+/-/- | + | + | + | 32 |
| **NOrMAL**: Accurate nucleosome positioning using a modified Gaussian mixture model. C++ code and executables are provided for download [128]. It is a command line tool designed to resolve overlapping nucleosomes and extract extra information ("fuzziness", probability, etc.) of nucleosome placement. Newer software called **PuFFIN** developed by the same authors is claimed to outperform **NOrMAL**. http://www.cs.ucr.edu/~polishka/ | +/-/-/- | - | + | + | 17 |
| **PING and PING 2.0**: Probabilistic inference for nucleosome positioning with MNase-based or sonicated short-read data. An R package for nucleosome peak calling integrated in the Bioconductor [108, 129]. The authors say that PING compares favorably to **NPS** and **TemplateFilter** in scalability, accuracy and robustness. http://www.bioconductor.org/packages/release/bioc/html/PING.html | -/+/-/- | - | + | + | 13 |
| **BINOCh**: Binding Inference from Nucleosome Occupancy Changes [141, 144]. This is a Python package, which allows identification of putative enhancers by comparing nucleosome occupancy in two cell conditions and analyzing DNA motifs near nucleosome centres and | +/-/-/- | - | - | + | 12 |

| Tool | | | | | |
|---|---|---|---|---|---|
| edges. It requires as input sorted BED files and relies for peak calling on the software **NPC** developed by the same group. http://liulab.dfci.harvard.edu/BINOCh/ | | | | | |
| **MLM**: A Multi-Layer Method to analyze microarray nucleosome positioning data. A Matlab code is available for download [46]. http://www.math.unipa.it/pinello/mlm/ | -/-/+/- | + | + | - | 12 |
| **NucPosSimulator**: Deriving non-overlapping nucleosome configurations from MNase-seq data [140]. It utilizes a Monte Carlo (MC) approach to determine the most probable nucleosome position in overlapping and ambiguous DNA reads from high through-put sequencing experiments. In contrast to peak-calling procedures NucPosSimulator probes many possible solutions, and can apply a Simulated Annealing scheme, a heuristic optimization method, which finds an optimal solution for complex positioning problems. http://bioinformatics.fh-stralsund.de/nucpos/ | -/+/-/+ | - | + | + | 10 |
| **NucDe**: Mapping nucleosome-linker boundaries [130]. This is an R package mapping nucleosome-linker boundaries from both MNase-Chip and MNase-Seq data using a non-homogeneous hidden-state model based on first order differences of experimental data along genomic coordinates. http://www.stat.wisc.edu/~keles/Software/demo_Nucde.pdf | -/+/-/- | + | + | - | 9 |
| **NucleoFinder**: A statistical approach for the detection of nucleosome positions [131]. An R package, which addresses both the positional heterogeneity across cells and experimental biases. https://sites.google.com/site/beckerjeremie/home/nucleofinder | -/+/-/- | - | + | + | 8 |
| **ChIPseqR**: Analysis of ChIP-Seq experiments using "R"; included in the Bioconductor R package [132]. ChIPseqR takes as input mapped reads and outputs nucleosome centres and their scores. It allows producing basic statistical graphs using standard R functions. http://www.bioconductor.org/packages/release/bioc/html/ChIPseqR.html | -/+/-/- | - | + | - | 7 |
| **DiNuP**: A systematic approach to identify regions of differential nucleosome positioning [138]. DiNuP compares the nucleosome profiles generated by high-throughput sequencing between different conditions. It provides a statistical P-value for each identified differential regions and empirically estimates the False Discovery Rate (FDR) as a cutoff when two samples have different sequencing depths and differentiate differential regions from the background noise. http://www.tongji.edu.cn/~zhanglab/DiNuP/ | +/-/-/- | - | - | ? | 7 |
| **NSeq**: a multithreaded Java application for finding positioned nucleosomes from sequencing data [133]. NSeq includes a user-friendly graphical interface written in Java. It computes FDRs for candidate nucleosomes from Monte Carlo (MC) simulations, plots nucleosome coverage and centers, and exploits the availability of multiple processor cores by parallelizing its computations. NSeq analyzes alignment data in BAM, SAM, or BED format. It assumes that the data are single-end. https://github.com/songlab/NSeq | -/-/-/+ | - | + | - | 7 |
| **Skyline** nucleosome browser: a web-based application for the identification of nucleosome peaks over the genome [142]. http://chromatin.unl.edu/cgi-bin/skyline.cgi | Web | - | - | + | 6 |

| Tool | | | | | |
|---|---|---|---|---|---|
| **NucHunter**: Inferring nucleosome positions with their histone mark annotation from ChIP-seq data [134]. It uses data from histone ChIP-seq experiments to infer positioned nucleosomes, and can predict positioned nucleosomes from one or multiple BAM files, e.g. taking into account a control experiment. http://epigen.molgen.mpg.de/nuchunter/ | -/-/-/+ | - | + | + | 5 |
| **Perl scripts** to analyze paired-end MNase-seq experiments [145]. The authors have listed the code in supplementary materials of their publication, which is useful for other developers. http://nar.oxfordjournals.org/content/suppl/2011/07/24/gkr643.DC1/Cole_Supp_Info.pdf | +/-/-/- | - | + | - | 5 |
| **iNPS**: The authors developed an improved version of the **NPC** nucleosome peak calling algorithm, which they claim to outperform the latter [135]. http://www.picb.ac.cn/hanlab/iNPS.html | +/-/-/- | - | + | + | 3 |
| **NUCwave**: Nucleosome occupancy maps from MNase-seq, ChIP-seq and CC-seq [139]. It is a Python package which generates nucleosome occupancy maps from MNase-seq, ChIP-seq and chemical cleavage (CC-seq), both for single-end and paired-end reads. It requires as input files in a Bowtie output format. http://nucleosome.usal.es/nucwave/ | +/-/-/- | - | - | + | 3 |
| **PuFFIN**: A parameter-free method to build genome-wide nucleosome maps from paired-end sequencing data [136]. PuFFIN is a command line tool for accurate placing of the nucleosomes based on the pair-end reads. It was designed to place non-overlapping nucleosomes using extra length information present in pair-end data-sets. PuFFIN is written in Python, and released in 2014. It outperforms **NOrMAL** previously released by the same authors, and is claimed by the authors to outperform also **NSeq**, **NPS** and **Template Filtering**. It returns nucleosome positions, the width of the peak, confidence score and fuzziness. http://www.cs.ucr.edu/~polishka/indexPuffin.html | +/-/-/- | - | + | + | 2 |
| **NucleoATAC:** A Python package for calling nucleosomes using ATAC-Seq data [40]. Requires as input sorted aligned paired-end reads in BAM format, FASTA file with genome reference and sorted bed file with non-overlapping regions for which nucleosome analysis is to be performed. These regions will generally be broad open-chromatin regions. Outputs nucleosome calls and occupancy. https://github.com/GreenleafLab/NucleoATAC | +/-/-/- | - | + | + | - |

## 3. Computational tools to predict nucleosome positions from the DNA sequence.

While nucleosome positioning is cell-type dependent, the DNA sequence still plays an important role in directing preferential histone octamer assembly. These methods are based on different ideas: either the physical considerations of DNA bendability, or bioinformatics analysis based on experimentally known nucleosome occupancies in specific cell types, or a combination of these approaches (Table 3). Correspondingly, existing web servers can be roughly split into four classes: The first class is based on purely bioinformatics considerations [49, 146-157], e.g. counting oscillatory dinucleotide distributions as pioneered in 1970s-80s [158], or moving the scanning window along the DNA and comparing the motif to the "ideal" nucleosome positioning motif [146-148], or introducing shorter nucleosome positioning nucleotide words, or a combination thereof. The second class is a hybrid of bioinformatics and biophysics [35, 52, 159-162], based on the algorithms which use dynamic programming to calculate allowed configurations of non-overlapping nucleosomes and variations of the Percus equation to assign the nucleosome formation energies learned from experimental nucleosome occupancy profiles. More details about the application of dynamic programming and the Percus equation for nucleosome positioning can be found elsewhere [27, 159, 163-167]. The third class is using correlations between empirical DNA characteristics, such as A-philicity, base stacking, B-DNA twist, bendability, bending stiffness, DNA denaturation energy, Z-DNA potential, etc, without knowing the underlying molecular details [168]. The fourth class of web servers goes further to the physics and calculates DNA bendability (which determines its affinity for the histone octamer) from first principles, assigning energetic-based scores to the dinucleotides, repetitions of dinucleotides, and in some cases to longer nucleotide words [168-174]. The latter approach is less dependent on learning nucleosome positioning rules from high-throughput sequencing, and usually uses for the parameterization either available crystal structures or high-throughput computer simulations [170, 175, 176]. The advantage of this approach is the possibility to include in the consideration covalent modifications of DNA and histones and even nucleosomes with partially unwrapped DNA [177, 178]. It is beyond the scope of the current review to compare the efficiency of these methods. A description of the main features of each algorithm is included in Table 3. Most original articles introducing new software cited in Table 3 did perform benchmarking against several existing algorithms. However, it is worth to note that since nucleosome positioning is regulated by several mechanisms, there are probably several classes of nucleosomes, each characterised by its preferred DNA pattern. Therefore, different algorithms predicting nucleosome positioning might be complementary rather than mutually exclusive. For example, we recently showed that while the algorithm of Segal and co-authors [35, 52, 159] predicts a single nucleosome enrichment peak at binding sites of a transcription factor CTCF, the algorithm of Trifonov and colleagues [32, 155, 179] intriguingly finds another class of "strong nucleosomes" regularly positioned around CTCF sites [26, 180] (in both cases calculations were performed in the absence of CTCF, not taking into account CTCF/nucleosome competition, which is yet another story [26]). Thus, there are many interesting biological processes determining nucleosome positioning which we still do not understand completely, and new algorithms are destined to appear. Nevertheless, readers interested to compare the performance of some of these methods are referred to recent reviews, e.g. [181].

**Table 3. Computational tools to predict nucleosome positioning (sorted alphabetically).**

| Description | Online / local installation | Di(tri)nucleotide periodicity / specific motifs / empirical feature |
|---|---|---|
| **FineStr**: Single-base-resolution nucleosome mapping server [146-148]. The analysis is performed using the probe based on the 117-bp DNA bendability matrix derived from *C. elegans*. The authors suggested the universality of this pattern for other species. http://www.cs.bgu.ac.il/~nucleom/ | +/- | +/+/+/- |
| **ICM Web**: ICM Web allows users to assess nucleosome stability and fold any sequence of DNA into a 3D model of chromatin [169, 182]. The model is displayed in the visual browser JSmol or can be downloaded. ICM takes a DNA sequence and generates (i) a nucleosome energy level diagram, (ii) coarse-grained representations of free DNA and chromatin and (iii) plots of the helical parameters (Tilt, Roll, Twist, Shift, Slide and Rise) as a function of position. http://dna.engr.latech.edu/icm-du | +/- | -/-/-/+ |
| **iNuc-PhysChem**: Identifying nucleosomal or linker sequences from physicochemical properties [168]. The algorithm identifies nucleosomal sequences by incorporating twelve physicochemical properties defined elsewhere, such as A-philicity, base stacking, B-DNA twist, bendability, bending stiffness, DNA denaturation energy, Z-DNA potential. The model was trained on data from H. sapiens, C. elegans and D. melanogaster. http://lin.uestc.edu.cn/server/iNuc-PhysChem | +/+ | +/-/+/+ |
| **iNuc-PseKNC**: A sequence-based predictor for nucleosome positioning in genomes with pseudo k-tuple nucleotide composition [151]. This is another software package from the developers of **iNuc-PhysChem**. Here, the samples of DNA sequences were formulated using six basic DNA local structural properties trained on datasets from *H. sapiens, C. elegans* and *D. melanogaster*. http://lin.uestc.edu.cn/server/iNuc-PseKNC | +/- | +/-/+/+ |
| **Mapping_CC**: Displays the nucleosome predictions based on the DNA dinucleotide correlation pattern. This algorithm was initially associated with one of the first high-throughput genome-wide nucleosome maps in Yeast [49]. | -/+ | +/+/-/- |
| An updated version is available at http://nuclbrowser.ucoz.com/load/ | | |
| **MOSAICS**: Methodologies for Optimization and Sampling in Computational Studies [170]. Perl scripts and a precompiled package to perform training-free atomistic prediction of nucleosome occupancy based on all-atom force field calculations. The effect of DNA methylation can be taken into account. http://www.cs.ox.ac.uk/mosaics/nucleosome/nucleosome.html | -/+ | +/-/-/+ |
| **NucEnerGen**: Nucleosome energetics predictions based on high throughput sequencing [160]. It utilizes dynamic programming to calculate allowed nucleosome configurations and the Percus equation to infer sequence-dependent energies from the experimental occupancy profiles. http://nucleosome.rutgers.edu/nucenergen/ | -/+ | +/+/+/- |
| **nuMap**: A web application implementing the YR and W/S schemes to predict nucleosome positioning [171-173]. The methodology is based on the sequence-dependent anisotropic bending, which dictates how DNA is wrapped around a histone octamer. This application allows users to specify a number of options such as schemes and parameters for threading calculation and provides multiple layout formats. http://numap.rit.edu/app/dna/index.xhtml | +/- | +/+/-/+ |
| **NuPoP**: Nucleosome Positioning Prediction Engine [161, 162]. NuPoP is built upon a duration hidden Markov model, in which the linker DNA length is explicitly modeled. NuPoP outputs the Viterbi prediction, nucleosome occupancy score (from backward and forward algorithms) and nucleosome affinity score. NuPoP has three formats including a web server prediction engine, a stand-alone Fortran program, and an R package. The latter two can predict nucleosome positioning for a DNA sequence of any length. http://nucleosome.stats.northwestern.edu | +/+ | +/+/-/- |
| **Nu-OSCAR**: Nucleosome-Occupancy Study for Cis-elements Accurate Recognition. It is devoted to identifying binding sites of known transcription factors, which further incorporates nucleosome occupancy around sites on promoter regions. The derivation of the algorithm is based on a biophysical view of interactions between protein factors and nucleosome DNA. http://bioinfo.au.tsinghua.edu.cn/nu_oscar/oscar.html | +/- | +/+/-/- |
| **nuScore**: A nucleosome-positioning score calculator based on the DNA curvature properties [174]. This software allows an important type of analysis, where a user enters many sequences to calculate the average nucleosome energy profile. http://compbio.med.harvard.edu/nuScore | +/- | +/+/-/+ |
| **N-score**: MATLAB and Python codes using a wavelet analysis based model for predicting nucleosome positions from DNA sequence [149]. http://bcb.dfci.harvard.edu/~gcyuan/software.html | -/+ | +/+/-/- |
| **NXSensor**: Prediction of nucleosome-excluding sequences based on DNA bending properties [150]. It takes as input DNA sequences in FASTA format, and outputs nucleosome-excluding or nucleosome favouring segments. http://www.sfu.ca/~ibajic/NXSensor/ | +/- | +/+/-/+ |
| **Online nucleosomes position prediction by genomic sequence (Segal Lab)** [35, 52, 159]. Although it has no specific name, this is one of the most popular tools in this class, realized as a web server (allows analyzing a limited number of DNA sequences), and a stand-alone application which can be installed on a local cluster. It allows calculating nucleosome occupancy or nucleosome start site probability profiles of non-overlapping nucleosomes; alternatively, it is possible to calculate the net nucleosome formation energy profile. It uses machine learning for energy assignment based on the training | +/+ | +/+/+/- |
| datasets and dynamic programming to sample nucleosome configurations (similar to **NucEnerGen**, **NuPoP** and the algorithm of van Noort and co-authors**)**. http://genie.weizmann.ac.il/software/nucleo_prediction.html | | |
| **Online nucleosome position prediction (van Noort and co-authors)** [163]. This algorithm is based on dinucleotide distributions, but unlike other methods based on dinucleotide distributions it does not use machine learning and accounts only for the dinucleotide periodicity. In addition, this method uses dynamic programming to account for size exclusion and the Percus equation to assign nucleosome affinities (similar to **NucEnerGen**, **NuPoP** and the algorithm of Segal and co-authors mentioned above**)**. http://bio.physics.leidenuniv.nl/~noort/cgi-bin/nup3_st.py | +/- | +/+/-/- |
| **Phase:** A web server for prediction of the nucleosome formation probability based on (i) the 10-11 bp periodicities of dinucleotides and (ii) the typical pattern "linker - nucleosome - linker" defined by the authors [156]. http://wwwmgs.bionet.nsc.ru/mgs/programs/phase/ | +/- | +/+/-/- |
| **RECON**: A web server for prediction of the nucleosome formation potential learned from dinucleotide frequencies distribution for nucleosome positioning sequences [153, 154]. http://wwwmgs.bionet.nsc.ru/mgs/programs/recon/ | +/- | +/+/-/- |
| **SymCurv**: A program for nucleosome positioning prediction [152]. It calculates the curvature of the DNA sequence and uses a greedy algorithm to parse the sequence in nucleosome-bound and nucleosome-free segments. http://genome.crg.es/SymCurv/documentation.html | -/+ | -/-/-/+ |
| **Strong nucleosomes**: Based on a recent discovery of strong nucleosome positioning sequences which are visually seen as regular arrays in genomic sequence [32, 155, 179], the program from Trifonov's lab is finding a specific class of strongly positioned nucleosomes of the RR/YY and TA periodic types [155]. http://strn-nuc.haifa.ac.il:8080/mapping/home.jsf | +/- | +/+/-/- |

**Conclusions**

The hunting for nucleosomes in their natural genomic environment has been opened for years and is far from being completed. Experienced nucleosome hunters are aware of the typical habits of nucleosomes: their preference for certain DNA sequences, and the ability to hide themselves using many natural obstacles (e.g. being masked by non-histone proteins [183], bound by transcription factors [184] or being able to unwrap and re-wrap their DNA dynamically [178, 185]), as well as artificial complications such as sequencing biases and the effect of the chromatin digestion level [186-188]. Furthermore, some of the sequence preferences for nucleosome cleavage are not artificial, but can be also associated with natural processes such as apoptosis [189]. As mentioned in the introduction, this review was not intended to discuss fundamental questions of the field, neither it was possible to highlight many excellent works which did not aim to develop online tools or resources. We have just performed a systematic excurse to the nucleosome positioning resources and tools which are available online. This growing list has to be taken seriously and studied systematically while developing new nucleosome positioning tools or designing new high-throughput experiments. Since new tools and datasets are appearing at high rate, an updated version of this list will be maintained online.

**Acknowledgement**

Many authors of online tools listed above have helped by providing details about their resources. Three anonymous referees are acknowledged for interesting comments and encouragement to extend this work.

Key points:

- >50 genome-wide nucleosome-positioning datasets already exist for human, >70 datasets for mouse, and >150 datasets for lower eukaryotes.

- At least 40 computational tools for the analysis of experimental nucleosome positioning data and theoretical prediction of nucleosome positioning from DNA sequence are already available online.